\documentstyle[11pt,epsfig]{article}
\title{\bf Lorentz violation  and the speed of gravitational waves in brane-worlds}
\author{F. Ahmadi$^{1,2}$\thanks{email: fa-ahmadi@sbu.ac.ir},
S. Jalalzadeh$^1$\thanks{email: s-jalalzadeh@sbu.ac.ir} and H. R.
Sepangi$^{1}$\thanks{email: hr-sepangi@sbu.ac.ir}
\\ $^1${\small Department of Physics, Shahid Beheshti University, Evin,
Tehran 19839, Iran}\\$^2${\small Department of Physics, Shahid
Rajaee University, Lavizan, Tehran 16788, Iran}\\}
\begin{document}
\maketitle

\begin{abstract}
Lorentz violation in a brane-world scenario is presented and used
to obtain a relationship between the speed of gravitational waves
in the bulk and that on the brane. Lorentz violating effects would
manifest themselves in gravitational waves travelling with a
greater speed in the bulk than on the brane and this effect is
independent of the signature of the extra dimension.
\end{abstract}

\section{Introduction}
According to Einstein's principle  of relativity the maximal
velocity in our universe, $c$, is equal to the fundamental speed
in the Minkowski space-time. This equality is deeply embedded in
$4D$ Einstein field equations \cite{dam} and is confirmed
experimentally \cite{will}. Recent discussions on the alternative
claims that the constant entering space-time intervals, or the
speed of gravity, is different from $c$ can be found in
\cite{foma,Car}.

General relativity however, cannot describe gravity at high enough
energies and must be replaced by a quantum gravity theory. The
physics responsible for making a sensible quantum theory of
gravity is revealed only at the Planck scale. This cut-off scale
indicates the point where our old notion of nature breaks down. It
is therefore not inconceivable that one of the victims of this
break down is Lorentz invariance. It is thus interesting to test
the robustness of this symmetry at the highest energy scales
\cite{AP,CG,ABGG}. As usual in high energy physics, if the scale
characterizing new physics is too high then it cannot be reached
directly in collider experiments. In this case cosmology is the
only place where the effects of new physics can be indirectly
observed. Brane-world models offer a phenomenological way to test
some of the novel predictions and corrections to general
relativity that are implied by M-theory. Such models usually
assume that $c$ is a universal constant. For alternative
approaches where the speed of gravity can be different from $c$ in
a brane-world context, see \cite{Dvali,Deff,Damour}. It should be
emphasized that the assumption that the maximal velocity in the
bulk coincides with the speed of light on the brane must not be
taken for granted. In this regard, theories with two metric
tensors have been suggested with the associated two sets of ``null
cones,'' in the bulk and on the brane \cite{merab}. This is the
manifestation of violation of the bulk Lorentz invariance by the
brane solution. In some brane-world scenarios, the space-time
globally violates $4D$ Lorentz invariance, leading to apparent
violations of Lorentz invariance from the brane observer's point
of view due to bulk gravity effects. These effects are restricted
to the gravity sector of the effective theory while the well
measured Lorentz invariance of particle physics remains unaffected
in these scenarios \cite{csaba, burgess}. In a similar vein,
Lorentz invariance violation has been employed to shed some light
on the possibility of signals travelling along the extra dimension
outside our visible universe \cite{stocia}. In a different
approach a brane-world toy model has been introduced
\cite{rubakov} in an inflating $5D$ brane-world setup with
violation of $4D$ Lorentz invariance at an energy scale $k$.

In a previous paper \cite{AJS}, we studied Lorentz violation in a
brane-world scenario by introducing a vector field normal to the
brane along the extra dimension. This, in fact, was a
generalization of the theory suggested by Jacobson and Mattingly
\cite{JM,kostel} who had investigated Lorentz violation in a 4$D$
space-time. In this work, we study local Lorentz violation in the
same setting and address the question of the speed of propagation
of gravitational waves in the bulk as well as on the brane. We
find a relation between the maximal velocity in the bulk and the
speed of light on the brane. As it turns out, we find that Lorentz
violating effects would manifest themselves in gravitational waves
travelling with a speed different from light, providing a possible
detection mechanism in gravitational wave experiments. Another
interesting outcome is that the speed of light turns out to be
greater in the bulk than on the brane irrespective of the
signature of the extra dimension. This is interesting since in
brane theories without Lorentz violation, the only way the speed
of gravitational waves in the bulk could become greater than that
on the brane is when the extra dimension is taken to be
space-like, which is the usual assumption in brane theories.
\section{Field equations}
In the usual brane-world scenarios the space-time is identified
with a singular hypersurface (or 3-brane) embedded in a
five-dimensional bulk. Suppose now that $N^{A}$ is a given vector
field along the extra dimension, effectively making the associated
frame a preferred one. The theory we consider consists of the
vector field $N^{A}$ minimally coupled to gravity with an action
of the form\footnote{The upper case Latin indices take the values
0, 1, 2, 3 and 5 while the Greek indices run from 0 to 3.}.
\begin{equation}
S=\int d^{5}x \sqrt{- ^{(5)}g}\left[\frac{1}{2k_{5}^{2}}
\left(^{(5)}R+{\cal L}_{N}\right)+{\cal L}_{m}\right],\label{eq1}
\end{equation}
where $k_{5}^{2}$ is a constant introduced for dimensional
considerations, ${\cal L}_{N}$ is the vector field Lagrangian
density while ${\cal L}_{m}$ denotes the Lagrangian density for
all the other matter fields. In order to preserve general
covariance, $N^{A}$ is taken to be a dynamical field. The
Lagrangian density for the vector field is written as
\begin{equation} {\cal
L}_{N}={K^{AB}}_{CD}\nabla_{A}N^{C}\nabla_{B}N^{D}+\lambda
(N^{A}N_{A}-\epsilon),\label{eq2}
\end{equation}
where
\begin{equation}
{K^{AB}}_{CD}=-\beta_{1}g^{AB}g_{CD}-
\beta_{2}\delta^{A}_{C}\delta^{B}_{D}-\beta_{3}\delta^{A}_{D}\delta^{B}_{C}.\label{eq3}
\end{equation}
Here, $\beta_{i}$ are dimensionless parameters, $\epsilon=-1$ or
$\epsilon=1$ depending on whether the extra dimension is
space-like or time-like respectively and  $\lambda$ is a Lagrange
multiplier. This is a slight simplification of the theory
introduced by Jacobson and Mattingly \cite{JM}, where we have
neglected a quartic self-interacting term of the form
$(N^{A}\nabla_{A}N^{B})(N^{c}\nabla_{c}N_{B})$, as has been done
in \cite{CL}. We also define a current tensor ${J^{A}}_{C}$ via
\begin{equation}
{J^{A}}_{C}\equiv {K^{AB}}_{CD}\nabla_{B}N^{D}.\label{eq4}
\end{equation}
Note that the symmetry of ${K^{AB}}_{CD}$ means that
${J^{B}}_{D}={K^{AB}}_{CD}\nabla_{A}N^{C}$. With these definitions
the equation of motion obtained by varying the action with respect
to $ N^{A}$ is
\begin{equation}
\nabla_{A}J^{AB}=\lambda N^{B}.\label{eq5}
\end{equation}
The equation of motion for $\lambda$ enforces the fixed norm
constraint
\begin{equation}
g_{AB}N^{A}N^{B}=\epsilon, \hspace{1cm} \epsilon^{2}=1.\label{eq6}
\end{equation}
The choice $\epsilon=1$ ensures that the vector will be time-like.
Multiplying both sides of (5) by $N_{B}$ and using (\ref{eq6}), we
find
\begin{equation}
\lambda=\epsilon N_{B}\nabla_{A}J^{AB}.\label{eq7}
\end{equation}
One may also project into a subspace orthogonal to $N^{A}$ by
acting the projection tensor ${P^{C}}_{B}=-\epsilon
N^{C}N_{B}+\delta^{C}_{B}$ on equation (\ref{eq5}) to obtain
\begin{equation}
\nabla_{A}J^{AC}-\epsilon N^{C}N_{B}\nabla
_{A}J^{AB}=0.\label{eq8}
\end{equation}
This equation determines the dynamics of $N^{A}$, subject to the
fixed-norm constraint.

In taking the variation, it is important to distinguish the
variables that are independent. Our dynamical degrees of freedom
are the inverse metric $g^{AB}$ and the contravariant vector field
$N^{A}$. Hence, the Einstein equations in the presence of both the
matter and vector fields in bulk space are \cite{EJM}
\begin{eqnarray}
^{(5)}G_{AB}=^{(5)\!\!\!}R_{AB}-\frac{1}{2}g_{AB}
{^{(5)}\!}R=k_{5}^2{^{(5)}}T_{AB},\label{eq9}
\end{eqnarray}
where
\begin{equation}
 {^{(5)}}T_{AB}=^{(5)\!\!}T^{(m)}_{AB}+\frac{1}{k^{2}_{5}}T^{(N)}_{AB}.\label{eqa.1}
\end{equation}
Here, $T_{AB}^{(m)}$ is the five-dimensional energy-momentum
tensor and the stress-energy $T_{AB}^{(N)}$ is considered to have
the following form \cite{CL,TJDM}
\begin{eqnarray}
T^{(N)}_{AB}&=&2\beta_{1}\left(\nabla_{A}N^{C}\nabla_{B}N_{C}-
\nabla^C
N_{A}\nabla_{C}N_{B}\right)-2\left[\nabla_C\left(N_{(A}{J^C}_{B)}\right)
+\nabla_{C}\left(N^C J_{(AB)}\right)\right.\nonumber\\ &-& \left.
\nabla_{c}\left(N_{(A} {J_{B)}}^{C}\right)\right]+ 2\epsilon
N_D\nabla_C J^{CD}N_A N_B +g_{AB}{\cal L}_{N}.\label{eq10}
\end{eqnarray}
\section{Geometrical setup}
Let us now assume that the background manifold $\bar{v}_{4}$ is
isometrically embedded in a pseudo-Riemannian manifold $v_{5}$ by
the map ${\cal Y}:\bar{v}_4\rightarrow v_{5}$ such that
\begin{equation}
{\cal Y}^{A}_{,\mu}{\cal
Y}^{B}_{,\nu}g_{AB}=\bar{g}_{\mu\nu},\hspace{.5cm}{\cal
Y}^{A}_{,\mu}N^{B}g_{AB}=0,\hspace{.5cm}
N^{A}N^{B}g_{AB}=\epsilon. \label{eq1c}
\end{equation}
where $g_{AB}(\bar{g}_{\mu\nu})$ is the metric of the bulk (brane)
space $v_{5}(\bar{v}_{4})$ in arbitrary coordinate, ${{\cal
Y}^{A}}({{\cal X}^{\mu}})$ is the basis of the bulk (brane) and
$N^{A}$ is normal unite vector, orthogonal to the brane. Since
$N^{A}$ is a vector field along the extra dimension, we may write
\begin{equation}
N^{A}=\frac{\delta^{A}_{5}}{\phi}, \hspace{1cm}
N_{A}=(0,0,0,0,\epsilon\phi),\label{eq12}
\end{equation}
with $\phi$ being a scalar field. The perturbation  of
$\bar{v}_{4}$ with respect to a small positive parameter $y$ along
the normal unit vector  $N^{A}$ is given by
\begin{equation}
{\cal Z}^{A}(x^{\alpha},y)={\cal Y}^{A}+y \phi N^{A},\label{eq2c}
\end{equation}
where we have chosen $N^A$ to be orthogonal to the brane, thus
ensuring gauge independency and having the field $\phi$ dependent
on the local coordinates $x^\alpha$ only \cite{jal}. The
integrability conditions for the perturbed geometry are the Gauss
and Codazzi equations. To find the perturbed metric, $g_{\mu\nu}$,
we follow the same definitions as in the geometry of surfaces.
Consider the embedding equations of the perturbed geometry written
in the particular Gaussian frame defined by the embedded geometry
and the normal unit vector
\begin{equation}
{\cal Z}^{A}_{,\mu}{\cal
Z}^{B}_{,\nu}g_{AB}=g_{\mu\nu},\hspace{.5cm}{\cal
Z}^{A}_{,\mu}N^{B}g_{AB}=0,\hspace{.5cm}
N^{A}N^{B}g_{AB}=\epsilon. \label{eq5c}
\end{equation}
Using equations (\ref{eq2c}) and (\ref{eq5c}), we may express the
perturbed metric in the Gaussian frame defined by the embedding as
\begin{equation}
g_{\mu\nu}=\bar{g}_{\mu\nu}+2 y \phi(x^{\alpha})
\bar{K}_{\mu\nu}+y^{2} \phi^{2}(x^{\alpha})
\bar{g}^{\alpha\beta}\bar{K}_{\mu\alpha}\bar{K}_{\nu\beta},\label{eq6c}
\end{equation}
where $\bar{K}_{\mu\nu}$ is the extrinsic curvature of the
original brane and the metric of our space-time is obtained by
setting $y=0$ $(g_{\mu\nu}=\bar{g}_{\mu\nu})$. Using equations
(\ref{eq12}), (\ref{eq5c}) and (\ref{eq6c}), the metric of the
bulk is written as
\begin{equation}
dS^{2}=g_{\mu\nu}(x^{\alpha},y)dx^{\mu}dx^{\nu}+\epsilon
\phi^{2}(x^{\alpha})dy^{2},\label{eq11}
\end{equation}
where we have used signature $(+ - - - \epsilon)$ everywhere. The
Einstein equations (\ref{eq9}) contain the first and second
derivatives of the metric with respect to the extra coordinate.
These can be expressed in terms of geometrical tensors in $4D$.
The first partial derivatives can be written in terms of the
extrinsic curvature
\begin{equation}
K_{\mu\nu}=\frac{1}{2}{\cal
L}_{N}g_{\mu\nu}=\frac{1}{2\phi}\frac{\partial
g_{\mu\nu}}{\partial y}, \hspace{1cm}K_{A5}=0.\label{eq13}
\end{equation}
The second derivatives can be expressed in terms of the projection
$^{(5)}C_{\mu5\nu5}$ of the bulk Weyl tensor to 5$D$
\begin{equation}
^{(5)}C_{ABCD}=^{(5)\!\!\!}R_{ABCD}-\frac{2}{3}\left(^{(5)\!\!}R_{A[C}g_{D]B}-
^{(5)\!\!\!}R_{B[C}g_{D]A}\right)+\frac{1}{6}\left(^{(5)\!\!}R
g_{A[C}g_{D]B}\right).\label{eq14}
\end{equation}
 In the absence of off-diagonal terms $(g_{5\mu}=0)$ the dimensional
reduction of the five-dimensional equations is particularly simple
\cite{ponce}, \cite{JPL}. Thus, the field equations (\ref{eq9})
can be split up into three parts \cite{AJS}
\begin{eqnarray}
^{(4)}G_{\mu\nu}&=&\frac{2}{3}k_{5}^{2}\left[^{(5)}T_{\mu\nu}+\left(^{(5)}T^{5}_{5}-\frac{1}{4}
(^{(5)}T)\right)g_{\mu\nu}\right]\nonumber\\
&-&\epsilon\left(K_{\mu\alpha}K^{\alpha}_{\nu}-K
K_{\mu\nu}\right)+\frac{\epsilon}{2}g_{\mu\nu}\left(K_{\alpha\beta}K^{\alpha\beta}-K^{2}\right)-\epsilon
E_{\mu\nu}, \label{eq15}
\end{eqnarray}
\begin{equation}
\phi^{\mu}_{;\mu}=-\epsilon\frac{\partial K}{\partial y}-\phi
\left(\epsilon
K_{\alpha\beta}K^{\alpha\beta}+^{(5)\!\!}R^{5}_{5}\right),
\label{eq17}
\end{equation}
\begin{equation}
D_{\mu}(K^{\mu}_{\nu}-\delta^{\mu}_{\nu}K)=
k^{2}_{(5)}\frac{^{(5)}T_{5\nu}}{\phi}.\label{eq18}
\end{equation}
In the above expressions, $E_{\mu\nu}$  is the electric part of
the Weyl tensor and the covariant derivatives are calculated with
respect to $g_{\mu\nu}$, {\it i.e.} $D g_{\mu\nu}=0$.
\section{Brane world considerations}
To progress any further we need the Einstein field equations on
the brane. We therefore concentrate on deriving these equations in
this section by presenting a quick and brief review on how this
can be done. The reader may consult reference \cite{AJS} for a
detailed discussion.

With the brane-world scenario in mind, it is assumed that the
five-dimensional energy-momentum tensor has the form
\begin{equation}
^{(5)}T^{(m)}_{AB}=\Lambda_{5}g_{AB},\label{eq19}
\end{equation}
where $\Lambda_{5}$ is the cosmological constant in the bulk. Now,
using equation (\ref{eq10}) we may calculate $^{(5)}T_{\mu\nu}$,
$^{(5)}T^{5}_{5}$ and $^{(5)}T$, obtaining
\begin{eqnarray}
^{(5)}T_{\mu\nu}&=&\frac{1}{k^{2}_{5}}\left[-4(\beta_{1}+\beta_{3})K_{\mu\gamma}K^{\gamma}_{\nu}+2(\beta_{1}+\beta_{3})K
K_{\mu\nu}+\beta_{2}g_{\mu\nu}K^{2}+\frac{2(\beta_{1}+\beta_{3})}{\phi}K_{\mu\nu,5}\nonumber
\right.\\
&+& \left. \frac{2\beta
{2}}{\phi}g_{\mu\nu}K_{,5}-(\beta_{1}+\beta_{3})g_{\mu\nu}K_{\alpha\beta}
K^{\alpha\beta}+2\epsilon\beta_{1}\frac{\phi_{,\mu}\phi_{,\nu}}{\phi^{2}}-
\epsilon\beta_{1}g_{\mu\nu}\frac{\phi_{,\alpha}\phi_{,}^{\alpha}}{\phi^{2}}\right]+\Lambda_{5}g_{\mu\nu},
\label{eq20}\\
^{(5)}T^{5}_{5}&=&\frac{1}{k^{2}_{5}}\left[(\beta_{1}+\beta_{3})K_{\alpha\beta}K^{\alpha\beta}+\beta_{2}K^{2}+
2\epsilon\beta_{1}g^{\mu\nu}\frac{\phi_{;\nu\mu}}{\phi}
-\epsilon\beta_{1}\frac{\phi_{,\alpha}\phi_{,}^{\alpha}}{\phi^{2}}\right]+\Lambda_{5},\label{eq21}\\
^{(5)}T&=&\frac{1}{k^{2}_{5}}\left[-3(\beta_{1}+\beta_{3})K_{\alpha\beta}K^{\alpha\beta}+2(\beta_{1}+\beta_{3})K^{2}+
5\beta_{2}K^{2}+\frac{2}{\phi}(\beta_{1}+\beta_{3})K_{,5}+\frac{8}{\phi}\beta_{2}
K_{,5} \nonumber \right.\\
&-& \left.
3\epsilon\beta_{1}\frac{\phi_{,\alpha}\phi_{,}^{\alpha}}{\phi^{2}}+
2\epsilon\beta_{1}g^{\mu\nu}\frac{\phi_{;\mu\nu}}{\phi}\right]+5\Lambda_{5}.
\label{eq22}
\end{eqnarray}
Now, upon defining the following new set of parameters
\begin{eqnarray}
\alpha_{1}&=&2(\beta_{1}+\beta_{3}), \hspace{1cm}
\alpha_{2}=\frac{2\epsilon(\beta_{1}+\beta_{2}+\beta_{3})}{3-2\epsilon
(\beta_{1}+4\beta_{2}+\beta_{3})}, \nonumber\\
\\
\alpha_{3}&=&\frac{\alpha_{1}(3+\epsilon-2\beta_{2})}{6}-\beta_{2},\hspace{1cm}
\alpha_{4}=\frac{\alpha_{1}(6+\epsilon+\alpha_{1})}{6},\hspace{1cm}
\alpha_{5}=\epsilon\beta_{1}, \nonumber\label{eq25}
\end{eqnarray}
and using equations (\ref{eq20}), (\ref{eq21}) and (\ref{eq22}),
equation (\ref{eq15}) may be written as
\begin{eqnarray}
^{(4)}G_{\mu\nu}&=&\frac{k^{2}_{5}}{2}g_{\mu\nu}\Lambda_{5}-\frac{3(\epsilon
+\alpha_{1})}{(3+\alpha_{1})}(K_{\mu\gamma}K^{\gamma}_{\nu}-K
K_{\mu\nu})-\frac{3(\epsilon+\alpha_{3})}{2(3
+\alpha_{1})}g_{\mu\nu}K^{2} \nonumber \\
&+&\frac{3(\epsilon+\alpha_{4})}{2(3+\alpha_{1})}g_{\mu\nu}K_{\alpha\beta}K^{\alpha\beta}
+\left[\frac{\alpha_{1}(\alpha_{5}+\frac{1}{2})+
3\alpha_{5}}{(3+\alpha_{1})}\right]g_{\mu\nu}
\frac{\phi_{;\alpha}^{\alpha}}{\phi}-
\frac{\alpha_{5}(5+\alpha_{1})}{2(3+\alpha_{1})}
g_{\mu\nu}\frac{\phi_{,\alpha}\phi_{,}^{\alpha}}{\phi^{2}}
\nonumber \\
&-&\frac{2\alpha_{1}}{(3+\alpha_{1})}\frac{\phi_{;\nu\mu}}{\phi}+
\left[\frac{4\alpha_{5}}{(3+\alpha_{1})}\right]
\frac{\phi_{,\mu}\phi_{,\nu}}{\phi^{2}}-\frac{3(\epsilon+\alpha_{1})}{(3+\alpha_{1})}E_{\mu\nu}.
\label{eq26}
\end{eqnarray}
Note that $(3+\alpha_{1})$ is the coefficient of the
four-dimensional Einstein tensor as one may multiply both sides of
the above equation by this factor. It therefore provides a
relation among the extrinsic curvature, the electric part of the
Weyl tensor and scalar field $\phi$ when $\alpha_{1}=-3$. We take
$\alpha_{1}\neq-3$ thereafter. In the spirit of the brane world
scenario, we assume $ Z_{2}$ symmetry about our brane, considered
to be a hypersurface $\Sigma$ at $y=0$. Using $Z_{2}$ symmetry,
the Israel junction conditions are obtained as
\begin{equation}
K_{\mu\nu}|_{\Sigma^{+}}=-K_{\mu\nu}|_{\Sigma^{-}}=-\frac{\epsilon
k^{2}_{5}}{2(1+\epsilon
\alpha_{1})}\left[\tau_{\mu\nu}-\frac{1}{3}g_{\mu\nu}(1+\alpha_{2})\tau\right].
\label{eq35}
\end{equation}
To avoid unreal singularities in equations (\ref{eq26}) and
(\ref{eq35}), it would be convenient to take $\alpha_{1}<-3$
\cite{AJS}. The energy-momentum tensor $ \tau_{\mu\nu}$ represents
the total vacuum plus matter energy-momentum. It is usually
separated in two parts,
\begin{equation}
\tau_{\mu\nu}=\sigma g_{\mu\nu}+T_{\mu\nu}, \label{eq38}
\end{equation}
where $ \sigma$ is the tension of the brane in $ 5D $, which is
interpreted as the vacuum energy of the brane world and
$T_{\mu\nu}$ represents the energy-momentum tensor of ordinary
matter in $4D$. Using equations (\ref{eq35}) and (\ref{eq38}) and
defining the following set of parameters
\begin{eqnarray}
\alpha_{6}&=&\frac{\alpha_{1}(1-2\alpha_{2})-2\epsilon\alpha_{2}}{3},
\ \nonumber\\
\\
\alpha_{7}&=&\frac{(\epsilon+\alpha_{1})(\alpha_{2}+\alpha_{2}^{2})}{3}+\frac{(\alpha_{4}-3\epsilon-4\alpha_{3})
(\alpha_{2}+2\alpha_{2}^{2})}{9}-\frac{(\alpha_{3}+2\alpha_{4})}{18},
\nonumber\label{eq39}
\end{eqnarray}
we obtain the Einstein field equations with an effective
energy-momentum tensor in $4D$ as
\begin{eqnarray}
^{(4)}G_{\mu\nu}&=&\Lambda_{4}g_{\mu\nu}+8 \pi G
T_{\mu\nu}+k^{4}_{5}\Pi_{\mu\nu}-
\frac{3(\epsilon+\alpha_{1})}{(3+\alpha_{1})}E_{\mu\nu}+\left[\frac{\alpha_{1}(\alpha_{5}+\frac{1}{2})+
3\alpha_{5}}{(3+\alpha_{1})}\right]g_{\mu\nu}
\frac{\phi_{;\alpha}^{\alpha}}{\phi}\nonumber \\&-&
\frac{\alpha_{5}(5+\alpha_{1})}{2(3+\alpha_{1})}
g_{\mu\nu}\frac{\phi_{,\alpha}\phi_{,}^{\alpha}}{\phi^{2}}
-\frac{2\alpha_{1}}{(3+\alpha_{1})}\frac{\phi_{;\nu\mu}}{\phi}+
\left[\frac{4\alpha_{5}}{(3+\alpha_{1})}\right]
\frac{\phi_{,\mu}\phi_{,\nu}}{\phi^{2}} , \label{eq40}
\end{eqnarray}
where
\begin{equation}
\Lambda_{4}=\frac{k^{2}_{5}}{2}\Lambda_{5}+\left[\frac{-\epsilon
k_{5}^{4}+3k_{5}^{4}(-\alpha_{1}+4\alpha_{6}+16\alpha_{7}+2\alpha_{4})
}{4(3+\alpha_{1}) (1+\epsilon\alpha_{1})^{2}}\right]\sigma^{2},
\label{eq41}
\end{equation}
\begin{equation}
8 \pi G=\left[\frac{-2\epsilon
k_{5}^{4}+3k_{5}^{4}(-2\alpha_{1}+4\alpha_{6})}
{4(3+\alpha_{1})(1+\epsilon\alpha_{1})^{2}}\right]\sigma,
\label{eq42}
\end{equation}
and
\begin{eqnarray}
\Pi_{\mu\nu}&=&\frac{3}{4(3+\alpha_{1})(1+\epsilon\alpha_{1})^{2}}
\left[-(\epsilon+\alpha_{1})T_{\mu\gamma}T^{\gamma}_{\nu}+(\frac{\epsilon}{3}+\alpha_{6})T
T_{\mu\nu}\right. \nonumber\\
&-&\left.
\left(\frac{\epsilon}{6}-\alpha_{7}\right)g_{\mu\nu}T^{2}+\frac{(\epsilon+\alpha_{4})}{2}g_{\mu\nu}
T_{\alpha\beta}T^{\alpha\beta}\right]+\left[\frac{3(\alpha_{6}+8\alpha_{7}+\alpha_{4})}
{4(3+\alpha_{1})(1+\epsilon\alpha_{1})^{2}}\right]g_{\mu\nu}\sigma
T. \label{eq43}
\end{eqnarray}
All these $4D$ quantities have to be evaluated in the limit $
y\rightarrow 0^{+}$. They give a working definition of the
fundamental quantities $ \Lambda_{4}$ and $G$ and contain
higher-dimensional modifications to general relativity. As
expected, switching off the effects of Lorentz violation
$(\alpha_i=0)$ in these equations results in expressions one
usually obtains in the brane-worlds models.  In the next section,
we use the solutions of equation (\ref{eq40}) to obtain a relation
between the speed of light in the bulk and on the brane.
\section{Speed of gravitational waves}
Let us start by assuming a perfect fluid configuration on the
brane. The energy-momentum tensor is therefore written as
\begin{equation}
T_{\mu\nu}=(\rho+p)u_{\mu}u_{\nu}-pg_{\mu\nu}, \label{eq44}
\end{equation}
where $\textbf{u}$, $\rho$ and $p$ are the unit velocity, energy
density and pressure of the matter fluid respectively. We will
also assume a linear isothermal equation of state for the fluid
\begin{equation}
p=\gamma\rho,\hspace{10mm} 0\leq\gamma\leq 1. \label{eq45}
\end{equation}
The weak energy condition \cite{HE} imposes the restriction
$\rho\geq0$. In this paper we deal with non-tilted homogeneous
cosmological models on the brane, {\it i.e.} we are assuming that
the fluid velocity is orthogonal to the hypersurfaces  of
homogeneity. In the standard cosmological models, We can also
consider $\phi(x^{\alpha})=\phi(t)>0$ \cite{ponce}.

Next, we consider the metric for our 4D universe as
\begin{equation}
\bar{g}_{\mu\nu}=\mbox{diag}(c_{b}^{2},-a(t)^{2}\Upsilon_{ij}),\label{eq46}
\end{equation}
with coordinates $(t,x^{i})$ and the 3-metric $\Upsilon_{ij}$ on
the spatial slices of constant time. Now, using the Israel
junction conditions, we have
\begin{eqnarray}
\bar{K}_{00}=\frac{-\epsilon k_{5}^{2}\bar{g}_{00}}{2(1+\epsilon
\alpha_{1})}\left[\sigma+\rho-\frac{1}{3}(1+\alpha_{2})(4\sigma+(1-3\gamma)\rho)\right],
\ \nonumber\\
\\
\bar{K}_{ii}=\frac{-\epsilon k_{5}^{2}\bar{g}_{ii}}{2(1+\epsilon
\alpha_{1})}\left[\sigma-\gamma\rho-\frac{1}{3}(1+\alpha_{2})(4\sigma+(1-3\gamma)\rho)\right].
\nonumber\label{eq48}
\end{eqnarray}
Now, by substituting the above equations in equations
(\ref{eq6c}), the different $4D$ sections of the bulk in the
vicinity of the original brane have the metric
\begin{equation}
g_{\mu\nu}=\Omega ^{2}\mbox{diag} (D c_{b}^{2},
-a(t)^{2}\Upsilon_{ij}),\label{eq47}
\end{equation}
where
\begin{eqnarray}
\Omega^{2}=(1+ y \phi B)^{2}  , \hspace{1cm}D=\left[\frac{1+ y
\phi A}{1+ y \phi B }\right]^{2},\label{eq49}
\end{eqnarray}
with
\begin{eqnarray}
A=\frac{-\epsilon k_{5}^{2}}{2(1+\epsilon
\alpha_{1})}\left[\sigma+\rho-\frac{1}{3}(1+\alpha_{2})(4\sigma+(1-3\gamma)\rho)\right],\\
B=\frac{-\epsilon k_{5}^{2}}{2(1+\epsilon
\alpha_{1})}\left[\sigma-\gamma\rho-\frac{1}{3}(1+\alpha_{2})(4\sigma+(1-3\gamma)\rho)\right].
\end{eqnarray}
From (\ref{eq46}), we see that the constant $c_{b}$ represents the
speed of light on the original brane, whereas from (\ref{eq47})
the speed of propagation of gravitational waves in this model is
$D c_{b}^{2}$. Also, within the context of this model, we have
$\frac{-\epsilon k_{5}^{2}}{2(1+\epsilon \alpha_{1})}>0$, so that
irrespective of the signature of the extra dimension, $A$ is
always greater than $B$ and consequently $D>1$. This leads to
apparent violations of Lorentz invariance from the brane
observer's point of view due to the bulk gravity effects. Also,
this result confirms the existence of a preferred frame at a point
in space-time. Now, if the effects of Lorentz violations are
ignored $(\alpha_{1}=0)$, the maximal velocity in the bulk becomes
more than the speed of light on the brane only when the extra
dimension is space-like. It is worth noting that if the energy
density of the matter fluid on the brane is zero, we obtain $A=B$,
implying that the maximal velocity in the bulk will be the speed
of light on the brane.

There is an interesting analogy between the behavior of
gravitational waves wandering into the bulk from the brane and
electromagnetic waves crossing one medium into another with
different indexes of refraction. This is a reflection of Fermat's
principle when studying gravitational wave propagation, that is,
when such waves take advantage of their greater speed in the bulk
and travell from one point on the brane to another by taking a
path through the bulk space, thus achieving a shorter travel time
than an electromagnetic wave traversing the same points.
Therefore, gravitational waves travelling faster than light would
be a possibility. These faster than light signals, however, do not
violate causality since the apparent violation of causality from
the brane observer's point of view is due to the fact that the
region of causal contact is actually bigger than the region one
would naively expect from the ordinary propagation of light in an
expanding universe. Indeed, there is no closed timelike curves in
the $5D$ spacetime that would make the theory inconsistent. The
importance of these models is that even extremely small Lorentz
violating effects may be measured in gravity wave experiments. For
example, an astrophysical event such as a distant supernova might
generate gravitational waves which would reach future gravity wave
detectors before we actually ``see'' the event. It is therefore
probable that future gravitational wave experiments like LIGO,
VIRGO or LISA might discover this unique signature of the
existence of extra dimensions. If found, such evidence may
strongly influence future developments in elementary particle
physics, cosmology and astrophysics \cite{CCEJ}.
\section{Conclusions}
In this paper we have studied a brane-world scenario where the
idea of Lorentz violation was considered by specifying a preferred
frame through the introduction of a dynamical vector field normal
to our brane. Such a normal vector was, however, assumed to be
decoupled from the matter fields since such fields were assumed to
be confined to the brane. The Einstein field equations were
obtained on the brane using the SMS formalism \cite{SMS}, modified
by additional terms emanating from the presence of the vector
field. As we have stressed, our model breaks the $4D$ Lorentz
invariance in the gravitational sector. Particle physics will not
feel these effects, but gravitational waves are free to propagate
into the bulk and they will necessarily feel the effects of the
variation of the speed of light along the extra dimension. We have
also shown that within the framework of our model, irrespective of
the signature of the extra dimension, the speed of the propagation
of gravitational waves is always greater in the bulk than on the
brane.

\end{document}